\documentclass[aps,twocolumn,showpacs,superscriptaddress]{revtex4-1}  % for review and submission
\usepackage{dcolumn}   % needed for some tables
\usepackage{bm}        % for math
\usepackage{amssymb}   % for math
\usepackage[dvips]{graphicx} % for figures
\usepackage{amsfonts}
\usepackage{amscd}
\usepackage{amsmath}    % need for subequations
\usepackage{enumerate}
\usepackage{epsfig}
\usepackage{subfigure}
\usepackage{natbib}
\usepackage{hyperref}
\usepackage{epstopdf}
\usepackage{latexsym}
\usepackage{float}
\usepackage{color}
\usepackage{multirow}

\begin{document}

\title{High-efficiency and low-jitter Silicon single-photon avalanche diodes based on nanophotonic absorption enhancement}

\author{Jian Ma}
\thanks{These authors contributed equally to this work}
\affiliation{Department of Modern Physics and National Laboratory for Physical Sciences at Microscale, Shanghai Branch, University of Science and Technology of China, Hefei, Anhui 230026, China}
\affiliation{CAS Center for Excellence and Synergetic Innovation Center in Quantum Information and Quantum Physics, Shanghai Branch,  University of Science and Technology of China, Hefei, Anhui 230026, China}
\author{Ming Zhou}
\thanks{These authors contributed equally to this work}
\affiliation{Department of Electrical and Computer Engineering, University of Wisconsin, Madison, Wisconsin 53706, USA}
\author{Zongfu Yu}
\affiliation{Department of Electrical and Computer Engineering, University of Wisconsin, Madison, Wisconsin 53706, USA}
\author{Xiao Jiang}
\affiliation{Department of Modern Physics and National Laboratory for Physical Sciences at Microscale, Shanghai Branch, University of Science and Technology of China, Hefei, Anhui 230026, China}
\affiliation{CAS Center for Excellence and Synergetic Innovation Center in Quantum Information and Quantum Physics, Shanghai Branch,  University of Science and Technology of China, Hefei, Anhui 230026, China}
\author{Yijie Huo}
\author{Kai Zang}
\affiliation{ Department of Electrical Engineering, Stanford University, Stanford, California 94305, United States}
\author{Jun Zhang}
\affiliation{Department of Modern Physics and National Laboratory for Physical Sciences at Microscale, Shanghai Branch, University of Science and Technology of China, Hefei, Anhui 230026, China}
\affiliation{CAS Center for Excellence and Synergetic Innovation Center in Quantum Information and Quantum Physics, Shanghai Branch,  University of Science and Technology of China, Hefei, Anhui 230026, China}
\author{James S. Harris}
\affiliation{ Department of Electrical Engineering, Stanford University, Stanford, California 94305, United States}
\author{Ge Jin}
\affiliation{Department of Modern Physics and National Laboratory for Physical Sciences at Microscale, Shanghai Branch, University of Science and Technology of China, Hefei, Anhui 230026, China}
\affiliation{CAS Center for Excellence and Synergetic Innovation Center in Quantum Information and Quantum Physics, Shanghai Branch,  University of Science and Technology of China, Hefei, Anhui 230026, China}
\author{Qiang Zhang}
\author{Jian-Wei Pan}
\affiliation{Department of Modern Physics and National Laboratory for Physical Sciences at Microscale, Shanghai Branch, University of Science and Technology of China, Hefei, Anhui 230026, China}
\affiliation{CAS Center for Excellence and Synergetic Innovation Center in Quantum Information and Quantum Physics, Shanghai Branch, University of Science and Technology of China, Hefei, Anhui 230026, China}

\date{\today}

\begin{abstract}
Silicon single-photon avalanche diode (SPAD) is a core device for single-photon detection in the visible and the near-infrared range, and widely used in many applications. However, due to limits of the structure design and device fabrication for current silicon SPADs, the key parameters of detection efficiency and timing jitter are often forced to compromise. Here, we propose a nanostructured silicon SPAD, which achieves high detection efficiency with excellent timing jitter simultaneously over a broad spectral range. The optical and electric simulations show significant performance enhancement compared with conventional silicon SPAD devices. This nanostructured devices can be easily fabricated and thus well suited for practical applications.
\end{abstract}

\maketitle

\section{Introduction}

Silicon single-photon avalanche diode (SPAD) operating in Geiger mode has become a standard device to detect ultra-weak optical signal in many fields, such as astronomy \cite{nightingale1990new,dravins2000avalanche}, biology \cite{li1993single,becker2005advanced,moerner2003methods}, lidar \cite{spinhirne1993micro,albota2002three}, quantum optics \cite{weihs1998violation,bouwmeester1997experimental}, quantum information \cite{hadfield2009single}. Compared with photomultiplier tube \cite{PMT}, for single-photon detection, Si SPAD has higher detection efficiency, lower dark count rate, and does not require a high voltage operation.

Currently, there are two primary types of Si SPADs, $Slik^{TM}$ structure and thin depletion layer structure. The representative devices based on $Slik^{TM}$ are SPCMs produced by PerkinElmer (now Excelitas). The detection efficiency of a typical device is higher than $50\%$ for the spectral range from 600 nm to 800 nm and its peak efficiency can be around $70~\%$ \cite{SPCM-AQ}. This performance benefits from a thick depletion layer(20-25$\mu$m) \cite{cova1996avalanche}, which guarantees an adequate absorption of single photons. However, the thick depletion layer induces a large timing jitter, their typical time resolution full-width at half maximum (FWHM) is about 400 ps \cite{SPCM-AQ}.

In the thin depletion layer structure, the thickness of depletion region is around 1$\mu$m \cite{lacaita1989double}. its timing resolution can be reduced down to 30 ps FWHM at room temperature \cite{MPD}. However, the thin depletion layer may result in inadequate absorption of single photons. The representative devices based on the narrow depletion layer are PDM photon counting modules produced by MPD. The peak detection efficiency of these devices is blue-shifted compared with SPCMs and the detection efficiency at 800 nm is about $15~\%$ \cite{MPD}.

% Fig1---structure of SPAD
\begin{figure*}
  \centering
  \includegraphics[width=16cm]{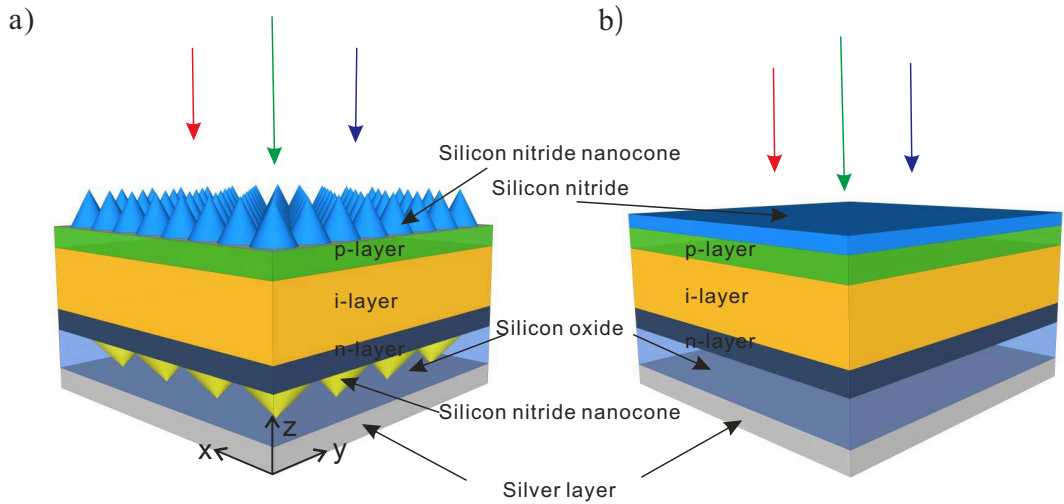}\\
  \caption{The structure of two silicon SPADs. (a)Nanostructured SPAD. There are silicon nitride nanocone gratings on both sides of the SPAD. For the upper gratings, the period is 400 nm, and the base diameter and the height of the nanocones are 400 nm and 800 nm, respectively. For the lower gratings, the period is 800 nm, and the base diameter and the height of the nanocones are 750 nm and 250 nm, respectively. On the bottom of the device there is a silver layer with a thickness of 200 nm. Between the silver layer and the lower gratings there is a silicon oxide spacer layer with a thickness of 2000 nm. (b) Conventional flat-film SPAD. Compared with the nanostructured SPAD, all the gratings are removed, and an additional antireflection layer of silicon nitride with a thickness of 100 nm is placed on the top instead. The dimensions of other layers are the same as the nanostructured SPAD.}
  \label{fig1}
\end{figure*}

One may ask an interesting question, whether it is possible to combine the excellent performance of high detection efficiency and low timing jitter in the same SPAD device. Solutions for this question have been previously investigated, e.g., using resonant cavity to enhance the efficiency without increasing the thickness \cite{ghioni2008resonant}. The cavity enhanced structure, however, induces a small spectral bandwidth of around a few nanometers, which severely limits its use in practice.
%Fig2 absorption--i-1um
\begin{figure}[!ht]
 \centering
  \includegraphics[width=8.2cm]{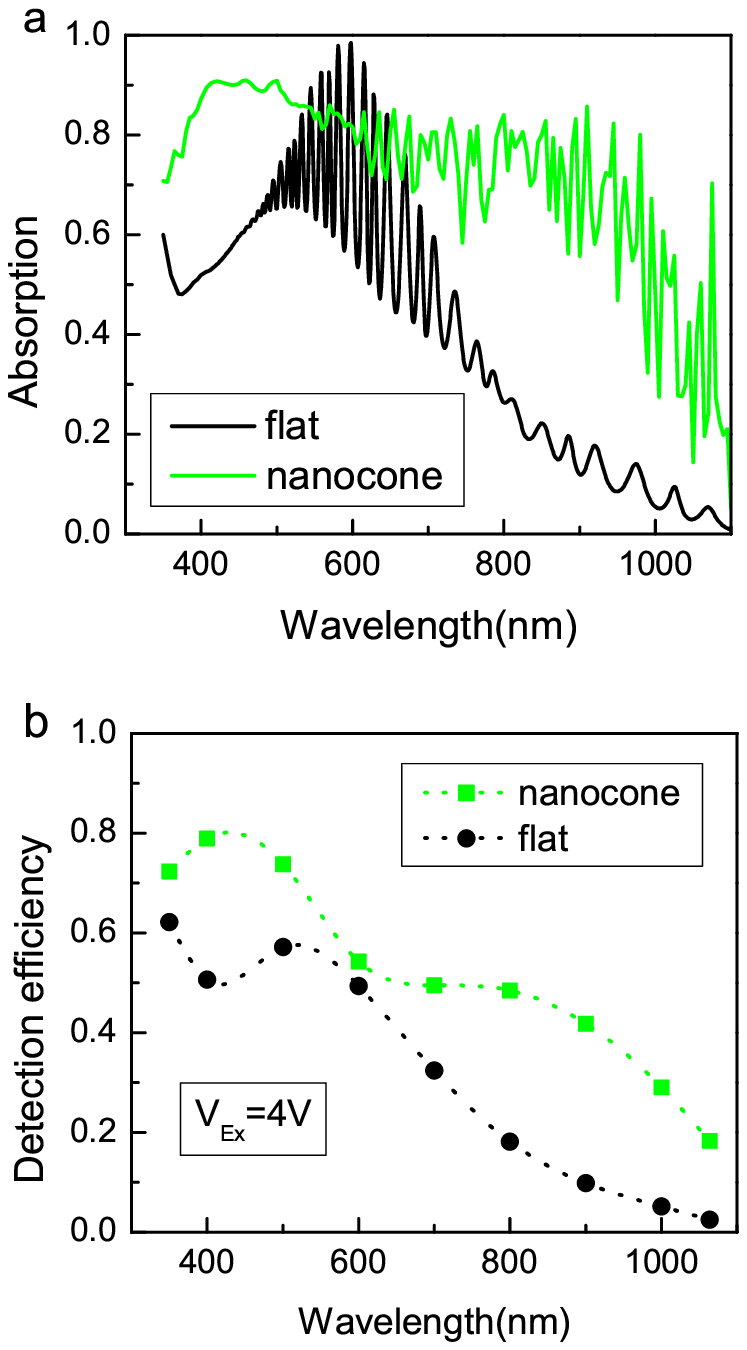}
  \caption{(a) The absorption spectrum of two SPADs. Black and green solid lines represent conventional thin film structure and nanocone structure, respectively. The nanostructured SPAD has an absorption efficiency higher than $60\%$ over the spectral range from 400 nm to 1000 nm. The difference between the two lines clearly shows the advantages of the nanostructure, particularly in the near-infrared region. (b) The detection efficiency as a function of wavelength with an excess bias voltage $V_{ex} = 4~V$. Here we calculate the detection efficiency at typical wavelength represented by solid dots and the dotted line between solid dots is just for guiding.}
  \label{fig2}
\end{figure}
In this paper, we first propose a nanostructured silicon SPAD to solve such problem. The nanostructured SPAD achieves high detection efficiency with excellent timing jitter over a broad spectral range. The enhancement mechanism is based on the same principles as light trapping enhancement used in solar cells \cite{yablonovitch1982statistical,yu2010fundamental,zhujia2008optical}. In the following sections, we first describe the new device structure and its enhancement mechanism. Then we provide optical and electric simulations to demonstrate the performance improvements. Finally, we briefly discuss about the feasibility of device fabrication.

%Second Paragraphe
\section{OPTICAL MODELING}
Photon detection efficiency $P_d$ depends on three parameters, i.e., photon absorption efficiency $P_a$, carrier collection efficiency $P_c$, and avalanche probability $P_b$. In this section, we analyze the light absorption in the semiconductor film of a SPAD, and show how to enhance the absorption efficiency using a nanostructure. For a thin film with a thickness of $d$, the single-pass absorption is given by
\begin{equation}
\label{eq1}
P_a=1-e^{-\alpha\cdot d},
\end{equation}
where $\alpha $ is the absorption coefficient. We assume that there is no reflection of light in the interfaces between the semiconductor film and the air. For silicon, the thickness $d$ has to be tens or even hundreds of micrometers in order to achieve an adequate absorption. Such thickness, however, severely limits other parameters of SPAD. We aim to enhance the photon absorption over a broad spectral range while reducing the required thickness. The same challenge arises in the development of solar cells, where the aim is to reduce the material cost of Silicon by using thinner film. For this purpose, light trapping has been developed to improve the light absorption in solar cells.  It allows a film to absorb much more light by using judiciously designed nanostructures to scatter light such that light propagates for a longer path in the film. A variety of structures has been proposed and implemented. It has been shown that the upper limit of light absorption by a film is \cite{campbell1987light,yablonovitch1982statistical,yu2012thermodynamic}

\begin{equation}
\label{eq2}
P_a^{u}=1-\frac{1}{1+4n^2\alpha d},
\end{equation}

where $n$ is the refractive index of the material. To see the effect of the absorption enhancement, we can take the limit of a very thin thickness such that $d/\alpha<<1$, the absorption calculated from Eq.\ref{eq2} is $4n^2$ times larger than that from Eq.\ref{eq1}. Based on light trapping, we design a highly efficient SPAD with enhanced light absorption over a broad spectral range.

The structure is based on a thin film with nanocone gratings coated on both sides as shown in Fig.\ref{fig1}a.
The top nanocone grating serves the purpose of broadband anti-reflection. The periodicity of the grating is 400 nm. The base diameter and the height of the nanocone is 400 nm and 800 nm, respectively. The choices of parameters are optimized for efficient anti-reflection. On the other hand, the bottom grating aims to scatter the light strongly toward the lateral direction. For this purpose, the periodicity of the grating is chosen as 800 nm,  which optimizes the scattering efficiency for near infrared light. The base diameter and the height of the nanocones are 750 nm and 250 nm, respectively. The combination of the effective anti-reflection and strong scattering is critically important in reaching the upper limit of the light absorption \cite{wang2012absorption}. We choose $Si_{3}N_{4}$ as the material for the nanocone gratings. Apart from the nanocone gratings, the SPAD is designed as a typical PIN structure as shown Fig.\ref{fig1}a, with p-layer of 300 nm thickness, i-layer of $1 ~\mu$m thickness, and n-layer of 300 nm thickness, both p-type region and n-type region is heavily doped. On the back side of the device there is a silver mirror with 200 nm thickness, and a silicon oxide layer with 2000 nm thickness is inserted between the lower gratings and the silver layer to reduce the absorption by the mirror. The optical absorption is simulated using S4 \cite{liu2012s}, which is based on the rigorous coupled wave analysis (RCWA) method \cite{li1997new,tikhodeev2002quasiguided}.

To verify the enhancement, a conventional SPAD with flat thin film as shown in Fig. \ref{fig1}b is also simulated for comparison. The upper gratings are replaced by an antireflection layer with 100 nm thickness, and the lower gratings are removed. The silicon PIN structure is the same as the nanostructured SPAD. There are also a silver mirror and a silicon oxide spacer on the back side of the film. For a fair comparison, a $Si_{3}N_{4}$ anti-reflection layer with a thickness of 100 nm is used on the top of the PIN. The light absorption in the PIN junction in both structures are calculated and shown in Fig.\ref{fig2}a. For the flat thin film structure, the absorption is around $60\%$ in the short wavelength range. However, in the near infrared (NIR) region, the absorption falls below $20\%$, indicating a poor efficiency for SPAD. In great contrast, the nanostructured SPAD mains high absorption around $80\%$ from visible to NIR regions.

\section{ELECTRIC MODELING}

Based on the simulation results of the absorption efficiency, we can further characterize the parameters of interest for single-photon detection, e.g., detection efficiency and timing jitter, through the simulations in Geiger mode.

For the SPADs as shown in Fig. \ref{fig1} there are three regions that can absorb incoming photons, i.e., the upper p-type region, the middle i-type region and the lower n-type region. In the different regions, the probabilities that the photo-generated carriers reach the depletion region, i.e., $P_c$, are also different. When a photon is absorbed in the depletion region, an avalanche is initiated immediately so that $P_c$ is close to 1. Since the depletion region is mainly in the intrinsic layer, the lifetime of photo-generated carriers is longer than the avalanche process such that the recombination effect can be ignored in this region \cite{gulinatti2009modeling}. When a photon is absorbed in the p-type region or the n-type region, due to the absence of high electric field the photo-generated minority carriers move randomly before reaching the depletion region. This process has an effect on detection efficiency and timing jitter as discussed previously \cite{gulinatti2009modeling}. In this paper, we use 1D random walk model to simulate such process. The diffusion coefficient in the simulations is taken from the literature \cite{spinelli1997physics}. As the boundary conditions of the model, we assume that random walk starts when the first electron-hole pair is generated in the neutral region, meanwhile because high quality of $SiO_2$$-Si$ interface, we ignore the recombination at the $SiO_2$$-Si$ interface.

As the last step for the evaluation of detection efficiency, we calculate the avalanche probability $P_b$, i.e., the probability that an electron-hole pair triggers a self-sustaining avalanche, by using the random path length (RPL) model \cite{tan2007theoretical}. In the RPL model, the random ionization path length of electrons and holes can be described by their probability density functions, which are used as inputs. For electrons, the probability density function is

\begin{equation}
\label{eq3}
  h_e(\xi)=\left\{
    \begin{array}{cc}
      0    \!\!\!&, \xi \leq d_e  \\
      \alpha_e exp[-\alpha _e \left( \xi -d_e \right)] \!\!\!\!&, \xi > d_e,
    \end{array}
  \right.
\end{equation}

%Fig3 jitter vs wavelength
\begin{figure}[!ht]
 \centering
  \includegraphics[width=8.2cm]{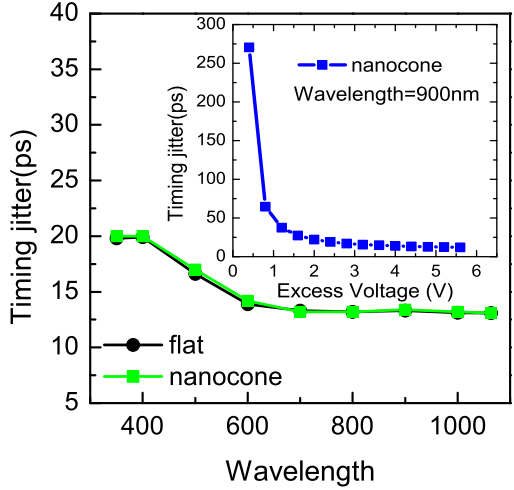}
  \caption{The timing jitter versus wavelength of two SPADs. Both SPADs are biased with same excess bias voltage $V_{ex} = 4~V$. The inset shows the excess voltage dependence of timing jitter at 900 nm for the nanocone SPAD.}
  \label{fig3}
\end{figure}

where $\alpha _e$ is the enabled ionization coefficient of electrons, and $d_e$ is the dead space length. $d_e=E_{the}/q\cdot f$,  where $E_{the}$ is the ionization threshold energy of electrons and $f$ is the electric field in the depletion region. The parameters of $\alpha _e$ and $d_e$ are calculated using the values of local ionization coefficient and $E_{the}$ taken from the literatures \cite{TRang,spinelli1996mean}. The probability density function of holes can be obtained similarly. In such a way, the RPL model can be effectively simulated using the Monte Carlo method \cite{Tan2007buildup}. For each trial, when an avalanche breakdown occurs, a detection event and threshold crossing time of avalanche current are recorded. After enough turns of trials, $P_b$ can obtained by calculating the ratio of the number of recorded events to the number of trials.

The results of detection efficiency simulation for the two SPADs are shown in Fig. \ref{fig2}b. Both SPADs are biased with same excess voltage $V_{ex} = 4~V$, the nanostructured SPAD exhibits much higher detection efficiency than the conventional SPAD at any wavelength. Particularly in the near-infrared range, the improvement of the detection efficiency is significant that is similar to the trend as shown in Fig. \ref{fig2}a.

%Fig4 jitter as a function of detection efficiency
\begin{figure}[htbp]
  %\centering
  \includegraphics[width=8.2cm]{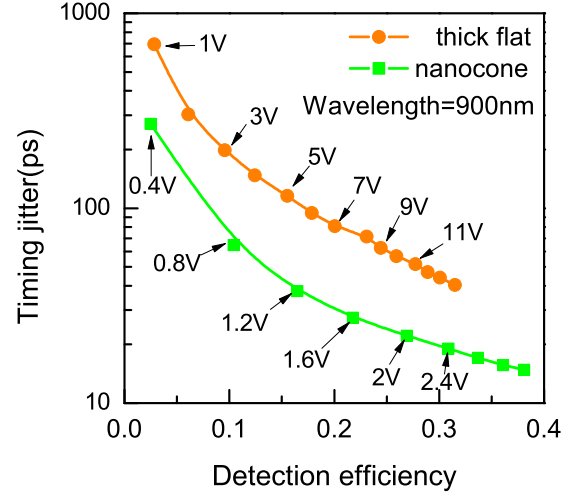}
  \caption{Timing jitter as a function of detection efficiency for nanostructured SPAD and flat-film SPAD with thick depletion region. Both timing jitter and detection efficiency change with the increasing of excess voltage, for some points the corresponded excess voltage have been indicated.}
  \label{fig4}
\end{figure}

Timing jitter is another key parameter of SPAD to characterize the time uncertainty between the photon absorption and avalanche detection. It can be calculated as

\begin{equation}
\sigma=\sqrt{<t_b^2>+<t_b>^2},
\end{equation}

where $t_b$ is defined as the time for an avalanche current reaching the threshold. For the SPADs, the timing jitter is mainly attributed to the time dispersion of the photo-generated carriers in the neutral region to reach the high field region, and the intrinsic randomness in avalanche process as well \cite{lacaita1995avalanche,spinelli1997physics}. In the simulations, we use the random walk method to evaluate the contribution of time dispersion in the neutral region. The intrinsic randomness in avalanche process includes the contributions from two parts, i.e.,  avalanche buildup process and propagation process \cite{spinelli1997physics}. Here, a low threshold current of 0.2 mA is used to detect avalanche occurrence, such that the avalanche process is still confined around the seed point. In this scenario, the avalanche buildup process dominates the intrinsic randomness \cite{spinelli1997physics}. The time uncertainty contribution due to the avalanche buildup process is extracted directly from the simulations of the RPL model.

\begin{figure}[!ht]
  \centering
  \includegraphics[width=8.2cm]{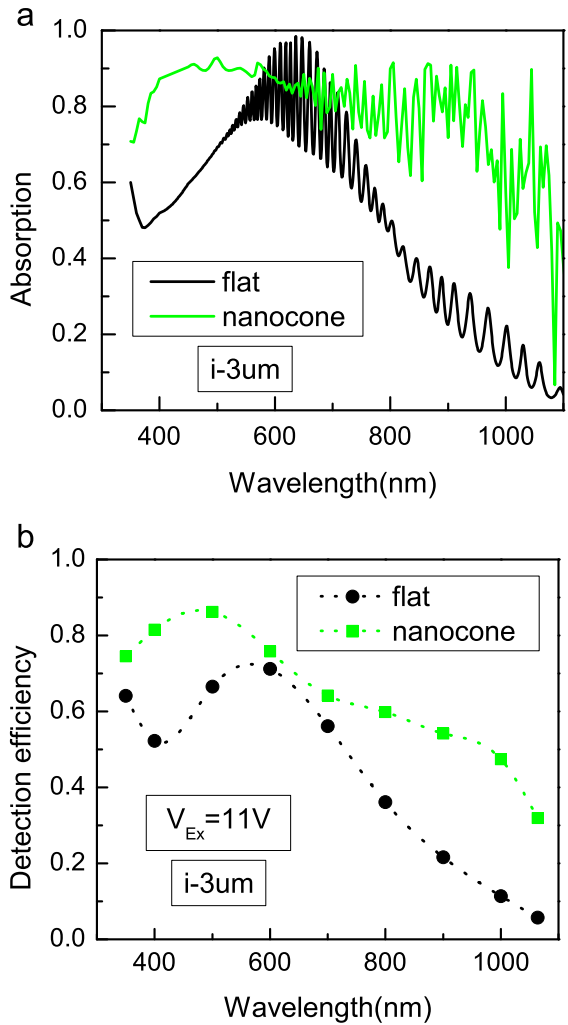}
  \caption{The thickness of the i-layer is increased to 3 $\mu$m. (a) Absorption spectrum for both structures when. Both structures have higher absorption efficiency. For the nanostructured SPAD (green solid line), the absorption efficiency is higher than $80\%$ over the spectral range from 400 nm to 1000 nm. (b) The detection efficiency as a function of wavelength with $V_{ex} = 11~V$. Here we calculate the detection efficiency at typical wavelength represented by solid dots and the dotted line between solid dots is just for guiding}
  \label{fig5}
\end{figure}

Fig. \ref{fig3} shows the timing jitter of the two SPADs as a function of wavelength with $V_{ex} = 4~V$. The overlap of two lines indicates that the timing jitter of the nanocone SPAD can be as low as the flat thin film SPAD. The inset in Fig. \ref{fig3} exhibits the timing jitter characteristics of the nanocone SPAD  decreases as the bias voltage increases.

Combining the simulation results of the detection efficiency and timing jitter, one can conclude that the nanostructured SPAD with a depletion width of 1 $\mu$m presents the advantages of high efficiency and low jitter simultaneously. Therefore, nanostructure is an effective approach to solve the coexistence problem of SPAD performance aforementioned.

\begin{figure}[!ht]
\includegraphics[width=8cm]{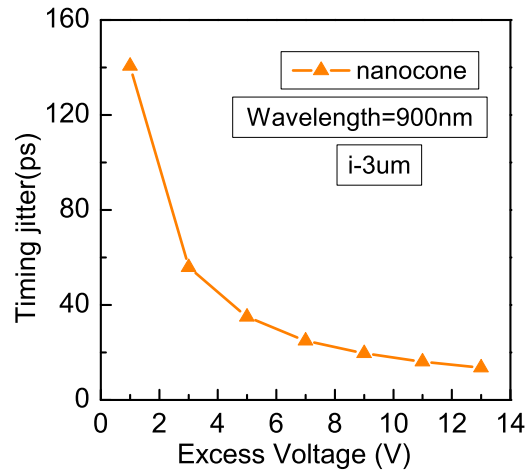}
\caption{Timing jitter as a function of excess voltage for the nanocone SPAD with thicker depletion region. }
\label{fig6}
\end{figure}

With the help of light absorption enhancement, the nanostructured SPAD achieves high absorption efficiency with a very thin depletion layer. By increasing the thickness of the depletion layer, such high absorption efficiency can also be achieved using the conventional flat-film SPAD. However, as the thickness of the depletion layer increases, the jitter performance of SPAD significantly decreases \cite{gulinatti2011i_red-enhanced-SPAD}. To verify such effect, we compare the jitter of the two SPADs at the same absorption efficiency. The depletion layer thickness of the flat-film SPAD is increased to keep the same absorption efficiency.

The timing jitter results are shown in Fig.~\ref{fig4}, at the wavelength of 900 nm. The large difference between the two lines in Fig.~\ref{fig4} shows a remarkable improvement on the timing jitter using the nanostructured SPAD. For instance, with a depletion layer thickness of $1~\mu$m, the nanostructured SPAD has a detection efficiency of $32~\%$. For the flat-film SPAD, in order to achieve the same efficiency the thickness of the depletion layer has to be increased from $1~\mu$m to $5.8~\mu$m. As a result of such thickness, the timing jitter increases with one order of magnitude.

By increasing the thickness of the depletion layer, we can also increase the absorption efficiency of the nanostructured SPAD. However, unlike the flat film structure, the nanostructured SPAD does not require an extremely thick depletion layer in order to obtain high absorption efficiency. Considering a nanostructured SPAD with a depletion layer of $3~\mu$m thickness, a p-layer of 100 nm thickness and a n-layer of 100 nm thickness, when nanostructured SPAD is biased with excess voltage $V_{ex} = 11~V$, $P_a$ is higher than $80~\%$ in the range from 400 nm to 1000 nm as shown in Fig.~\ref{fig5}a. $P_d$ reaches around $60~\%$ at the wavelength of 900 nm as shown in Fig.~\ref{fig5}b, which is much higher than that in the case of $1~\mu$m thickness as shown in Fig.~\ref{fig2}b.
The flat-film SPAD with the same layer parameters is also simulated for comparison. Without nanostructure the detection efficiency of flat-film SPAD drop to around $20~\%$ at the same wavelength. Similarly, we calculate the timing jitter performance of this thick nanostructured SPAD. In Fig.~\ref{fig6}, the timing jitter as a function of excess bias voltage is plot, and here we assume the wavelength of photon is 900 nm. With $V_{ex} = 11~V$, the timing jitter is only increased to 20 ps. On the other hand, for the flat film structure SPAD, the thickness of the depletion layer has to be $12 \mu$m to achieve a detection efficiency of $60\%$, indicating a much longer jitter time.

In addition to better device performance, this nanostructure is easy to be fabricated. Silicon nitride nanocone structure can be made by applying self-assembled nickel nano-particle and dry etching \cite{sahoo09}. The double-sided nanocone structure can be processed sequentially and one may even directly process on thin film silicon wafer \cite{wang13}. After nanocone etching, device backside can be coated with oxide and metal to form the final devices.

\section{CONCLUSIONS}

In summary, we have proposed and theoretically simulated a nanostructured SPAD that has the remarkable performance of high detection efficiency over a broad spectral range and low timing jitter at the same time. Our approach effectively solves the coexistence problem of high efficiency and low jitter for silicon SPADs.
Moreover, by optimizing the structure, the detection efficiency of the nanostructured SPAD could be as high as the conventional thick silicon SPAD, particularly in the near-infrared range, while maintaining a pretty low timing jitter. Such nanostructured SPAD is well suited for many practical applications requiring high efficiency and low timing jitter.

\section*{Acknowledgments}

Jian Ma and Ming Zhou contributed equally to this work. The authors would like to thank Xintao Bi, Jin Wang, Jinrong Wang for providing workstation for computation. This work has been supported by the National Fundamental Research Program (under Grant No. 2011CB921300, 2013CB336800), the National Natural Science Foundation of China, the Chinese Academy of Science.

% Bibliography

\end{document}